# Constructing a Unified Model of Community Formation in Community-Supported Agriculture: Insights from Consumer and Producer Pathways in Japan

Sota Takagi[1*], Miki Saijo[1], Takumi Ohashi[1]

[1]Institute of Science Tokyo, 2-12-1 Ookayama Meguro-ku, Tokyo 152-8550, Japan

* takagi.s.5b1d@m.isct.ac.jp

Abstract: Community Supported Agriculture (CSA) has been recognized globally as a promising framework that embeds agriculture within social relations, yet its diffusion remains limited in contexts such as Japan. Existing studies have largely focused on either consumer or producer participation in isolation, offering fragmented insights and leaving unexplored how their reciprocal processes jointly shape CSA communities. This study addresses this gap by integrating the trajectories of both groups into a comprehensive account of CSA community formation. Drawing on semi-structured interviews with ten CSA producers and ten consumers, we employed the Modified Grounded Theory Approach (M-GTA) to inductively theorize processes of participation and practice. The analysis showed that producers advance CSA through internal adjustments and sense-making to cope with uncertainties, while consumers are guided by life events, practical skills, and prior purchasing experiences toward participation. Synthesizing these insights, we propose a six-phase model of CSA community formation, dispersed interest, awareness, interest formation, motivation, practice, and co-creative continuation, that demonstrates how producers, consumers, and intermediaries interact across stages. The model highlights the pivotal role of key players in sustaining engagement and provides a new perspective for institutionalizing CSA as a durable component of sustainable food systems.

Keywords: Community-supported agriculture, Modified grounded theory approach, Community formation, Prosumption

1. Introduction

    In recent years, climate change, market volatility, and geopolitical tensions have heightened concerns about the resilience and sustainability of food systems (IPCC, 2019, 2023). Farmers face increasing uncertainties, including unstable yields, economic pressures, and social isolation (Wheeler et al., 2021; FAO, 2024), revealing the limitations of relying solely on producers to sustain agriculture. In this context, new frameworks are needed that embed support for farming within broader social relations and shared responsibilities across producers, consumers, and local communities. One promising framework that embodies this principle is Community-supported agriculture (CSA), which has been attracting growing attention worldwide (Feenstra, 2002; Egli et al., 2023).

    CSA is generally based on a system in which consumers become members, make advance

payments through annual contracts, and in return receive agricultural products on a regular basis. Because advance payments are not tied to actual harvest volumes, production risks arising from extreme weather, poor harvests, or disasters are shared between producers and consumers, allowing producers to stabilize their income. In addition, CSA is said to foster direct and continuous relationships between producers and consumers, thereby enhancing the locality and reliability of the food system(Medici et al., 2021; Egli et al., 2023). Despite these advantages, CSA remains a niche initiative, and its diffusion is still in its early stages in some countries and regions (Raftowicz et al., 2021; Middendorf and Rommel, 2024; Schmidt et al., 2024; Takagi et al., 2025).

Existing research has typically examined either consumers or producers in isolation. Prior studies have reviewed consumer participation factors and identified potential consumer segment (Birtalan et al., 2020; Huang et al., 2024; Takagi et al., 2024, 2025), while others have analyzed producers' motivations and barriers (Galt, 2013; Tay et al., 2024). However, few studies have traced the pathways by which consumers and producers, respectively, progress through changes in motivations, experiences, and interactions with key actors, and no research has integrated these trajectories into a unified model of CSA community formation. This gap is particularly critical in contexts where CSA remains emergent, as understanding the joint processes of participation and practice is essential for its broader diffusion.

The purpose of this study is to elucidate the participation and practice processes of both consumers and producers and to integrate these into a single phased model that explains how CSA communities are formed through their interactions. To achieve this, we conducted semi-structured interviews with CSA farmers and participating consumers, developed separate theoretical accounts of their respective CSA participation and practice processes, and then synthesized them into an integrated framework. By clarifying both sides of the process and demonstrating how they converge into a unified model, this study makes a distinct contribution to advancing theory and practice on CSA development.

2. Materials and method
2.1. Study design

This study employed a qualitative design to elucidate the processes through which producers and consumers engage in CSA and to integrate these into a comprehensive model of community formation. The research was situated within a social constructivist paradigm, which acknowledges that theories are co-constructed through interactions between researchers and participants, and that researchers' interpretations and reflexivity shape data collection and analysis. To analyze our interview data, we adopted the Modified Grounded Theory Approach (M-GTA) (Kinoshita, 2020). M-GTA is a methodological refinement developed in Japan that builds on GTA principles while providing systematic procedures, such as the use of analytic worksheets and constant comparison across cases, for generating substantive, practice-oriented theories. Rather than aiming for abstract universal explanations, M-GTA emphasizes theorizing grounded in participants' lived experiences and contextual practices. This made it particularly well suited to our research objective of clarifying the stepwise processes of participation and practice among CSA producers and consumers, and of integrating these trajectories into an explanatory model of community formation.

Separate analyses were first conducted for producers and consumers to identify the processes underlying CSA participation and practice from each perspective. These analyses generated categories capturing the dynamics specific to each group. Subsequently, the categories were systematically compared to identify points of convergence and divergence. Through iterative discussions among the research team, these insights were integrated into a phased model of CSA community formation. The procedures of concept generation, category building, and model integration are further detailed in the analytical procedure section.

2.2. Producer interview

To explore the processes and factors that led producers in Japan to adopt CSA, we conducted in-depth interviews with ten practitioners. Participants were recruited by the first author through direct outreach to producers using CSA platforms (Table 1). The selection criteria required that the producers had practiced CSA in Japan.

The first interview was conducted with Producer A, which served as a pilot to refine the interview protocol. Subsequent interviews were then conducted with Producers B through H. Two of the participating CSA farms were co-managed, resulting in joint interviews with two members each (G and G'; H and H').

**Table 1. List of research participants (Producers).**

| ID | years of practice | Interview date |
|---|---|---|
| A | 3 years | Aug. 27th 2024 |
| B | 3 years | Jan. 22nd 2025 |
| C | 3 years | Jan. 25th 2025 |
| D | 1 year | Jan. 29th 2025 |
| E | 3 years | Jan. 29th 2025 |
| F | 3 years | Jan. 31st 2025 |
| G, G' | 3 years | Feb. 1st 2025 |
| H, H' | 3 years | Feb. 2nd 2025 |

All interviews were conducted either in person or via Zoom, depending on the convenience of the participants. A semi-structured interview format was employed to elicit narratives related to the adoption of CSA, motivations, and perceived changes through practice. Participants who agreed to compensation were paid 1,500 JPY per hour for their time and expertise; no payment was made to those who declined compensation. With consent, all interviews were audio-recorded and transcribed verbatim for analysis.

2.3. Consumer interview

The consumer interviews were conducted to explore the processes and factors that led individuals in Japan to participate in CSA (Table 2). Interviewees were recruited through two primary methods: referrals from CSA producers listed in Table 1 and a monitor survey company that targeted individuals with an interest in CSA. This dual recruitment strategy allowed the inclusion of both consumers with direct CSA experience and those who expressed interest but may not have had deep engagement.

However, this approach also introduces several potential biases. First, participants referred by CSA producers may have close relationships with the producers and therefore exhibit more favorable evaluations of CSA. Second, those recruited through the monitor company may have limited CSA practices.

Table 2. List of research participants (Consumers).

| ID | Age | Job description | Recruitment methods | Interview date |
|---|---|---|---|---|
| A | 40s | Office worker | Monitor company | Apr. 17th 2025 |
| B | 30s | Public officer | Monitor company | Apr. 18th 2025 |
| C | 40s | Office worker | Monitor company | Apr. 20th 2025 |
| D | 50s | Office worker | Producer C | May 1st 2025 |
| E | 50s | Executive | Producer C | May 2nd 2025 |
| F | 50s | Office worker | Producer A | May 13th 2025 |
| G | 30s | Freelance | Producer A | May 16th 2025 |
| H | 50s | Teacher | Producer C | May 19th 2025 |
| I | 50s | Public officer | Producer C | May 21st 2025 |
| J | 60s | Freelance | Producer C | May 21st 2025 |

Each participant received monetary compensation of 1,500 JPY per hour in exchange for their time and the insights they provided regarding their CSA experiences. Compensation was not provided to those who explicitly declined it. All interviews were conducted online using Zoom.

A semi-structured interview format was employed, focusing on the participants' pathway to CSA involvement, their motivations, and the changes they experienced through their participation. With informed consent, all interviews were audio-recorded and transcribed verbatim for qualitative analysis.

In the case of Consumer G, due to technical issues, the audio recording for the first half of the interview was not captured. Therefore, detailed notes taken during the interview were used as supplementary data for analysis.

2.4. Modified-grounded approach

In this study, we adopted the M-GTA to analyze the interview data (Kinoshita, 2020). Data collection continued until theoretical saturation was reached when no new themes or insights emerged from the interviews (Breckenridge and Jones, 2009). To ensure falsifiability and transparency in presenting the results, we followed the steps below:

1. Defining the Analytical Theme and Focus

    Based on the research objectives, we defined analytical themes and focus subjects to clarify specific processes. For producers, the theme was "The practice process of CSA among producers in Japan"; for consumers, it was "The participation process in CSA among consumers in Japan". The corresponding focus subjects were "CSA-practicing producers" and "CSA-participating consumers".

2. Development of Analytical Worksheets

    The analytical worksheet template used in this study (Table 3) included columns for concept name, definition, variations (representative quotations), and theoretical memos.

**Table 3. A template of the analytical worksheet used in this study**

| Concept name | Definition | Theoretical memo | Variations |
|---|---|---|---|
|  |  |  | Interviewee's utterance regarding the extracted concept (ID_No., Line number) |

3. Concept Generation

    The first author extracted relevant portions from the transcripts in light of the analytical focus and entered them into the variation column. These examples were interpreted and recorded in the definition column, which was then condensed into concise expressions to serve as concept names, entered in the concept column. Interpretations or questions arising during the process were recorded in the theoretical memo column. Additional matching quotations were sought throughout the dataset, and when a sufficient number of variations had been collected, the concept was deemed valid.

4. Reverse Checking

   We verified that each concept and its definition corresponded appropriately to the full set of examples. We also considered similar cases not yet covered by the definition and revised the concept or definition accordingly to ensure close alignment with the data.

5. Category Formation

   Relationships among the generated concepts were examined to produce higher-order subcategories, which were then further grouped into overarching categories. These relationships were summarized in a relational diagram.

6. Creation of Result Diagrams and Storylines

   The relationships among categories, subcategories, and concepts were visualized in result diagrams. A corresponding storyline was then developed to narratively explain the findings.

7. Peer Debriefing

   To enhance the credibility and trustworthiness of the theorization, peer debriefing was conducted with co-authors after the initial grounded theory was generated (Janesick, 2015). The first author shared the relational diagrams and storyline, and in-person or online discussions were held. The fist and second author critically reviewed the findings, suggested revisions, and resolved any differences through consensus. This collaborative process strengthened the robustness of the theory and ensured that it accurately reflected the data.

M-GTA analysis commenced after all interviews were completed. Each transcript was analyzed sequentially with constant comparison, integrating new insights and refining concepts iteratively. Theoretical saturation was considered reached when no new concepts emerged and all concepts, subcategories, and categories were coherently interconnected. In this study, saturation was confirmed in the final interviews; had new concepts emerged, additional interviews would have been conducted to ensure comprehensive coverage.

2.5. Process of Integrating Consumer and Producer Accounts

After constructing separate theoretical accounts of consumers and producers using the M-GTA, we proceeded to integrate these accounts into a unified model. This integration was not a mechanical merging of categories but an iterative and reflexive process. As a guiding orientation, we drew on the phronetic iterative approach (Tracy and Hinrichs, 2017; Tracy, 2018), which emphasizes focusing, integrating, and refining insights through cycles of interpretation.

In addition, we followed the reflexive framework (Srivastava and Hopwood, 2009), using guiding questions such as: "What are the data telling me?", "What is it I want to know?", and "What is the dialectical relationship between what the data are telling me and what I want to know?" These questions provided a structure for systematically revisiting the full set of concepts, subcategories, categories, and interview excerpts in relation to the evolving focus on community formation.

Through this iterative process, we identified points of divergence and convergence that ultimately informed the phased representation of CSA community formation presented in the Results section. Repeated discussions with co-authors provided critical reflection, challenged initial interpretations, and ultimately enhanced the validity and stability of the final model.

2.6. Researcher characteristics and reflexivity

This study was conducted by three researchers with diverse academic backgrounds. The first author had participated in a CSA platform as a consumer for approximately one year, which provided a basic understanding of its mechanisms and operations. The author also visited farms to observe agricultural practices and living conditions and had prior interactions with some consumer participants. In addition, prior research conducted by the author on CSA participation factors and design-based engagement (Takagi et al., 2024, 2025) informed the analytical sensitivity toward relational dynamics, learning processes, and motivational shifts observed in this study. These experiences enriched contextual understanding but may also have influenced interpretations.

The second author was born and raised in Tokyo, where there is no agricultural environment nearby. However, through a consumer cooperative (co-op) platform, the author had approximately one year of experience purchasing agricultural products directly from specific farmers. The author eventually discontinued the service due to the inconvenience of receiving home deliveries. These experiences shaped the author's sensitivity to the practical constraints and everyday decision-making processes that urban consumers face when considering participation in CSA or similar schemes.

The third author, a transition design researcher, has previously employed the M-GTA to theorize producers' decision-making and adaptation in Japan's livestock sector (Ohashi et al., 2024). Drawing on this experience, the author contributed to integrating producer perspectives and ensuring methodological coherence in the present study. Prior participatory research on short value chains (Washio et al., 2025) and consumer segmentation (Washio et al., 2023) also informed the framing of community formation.

To mitigate such potential biases, the study adopted the M-GTA. Throughout the analytical process, diagrams and narrative descriptions were combined, and peer debriefing sessions within the research team were conducted to critically review findings, challenge interpretations, and make necessary revisions, thereby ensuring transparency and trustworthiness.

In the final stage of model integration, potential bias was further addressed by first developing consumer and producer processes separately, and then comparing them step by step to identify points of convergence and divergence before constructing the unified model. The integration was conducted through iterative team discussions, where alternative interpretations were critically examined, and consistency was ensured by repeatedly cross-checking the original data with analytic records. While researcher influence cannot be entirely eliminated within a social constructivist paradigm, these procedures enhanced reflexivity and minimized undue bias in theorizing the community formation process.

3. Results

3.1. CSA implementation process for producers

This section presents the results of the M-GTA analysis of the interview records with producer (Fig. 1). The analysis generated 48 initial concepts. These concepts were compared and grouped into 18 subcategories (Table A1), each representing a higher level of abstraction that synthesizes multiple related concepts. Subsequently, the subcategories, together with unclassified concepts, were further integrated and abstracted to yield 6 overarching categories (Table A2). This hierarchical structuring, moving from concepts to subcategories and then to categories, provides an analytical framework for systematically understanding the factors that influence producers' CSA practices.

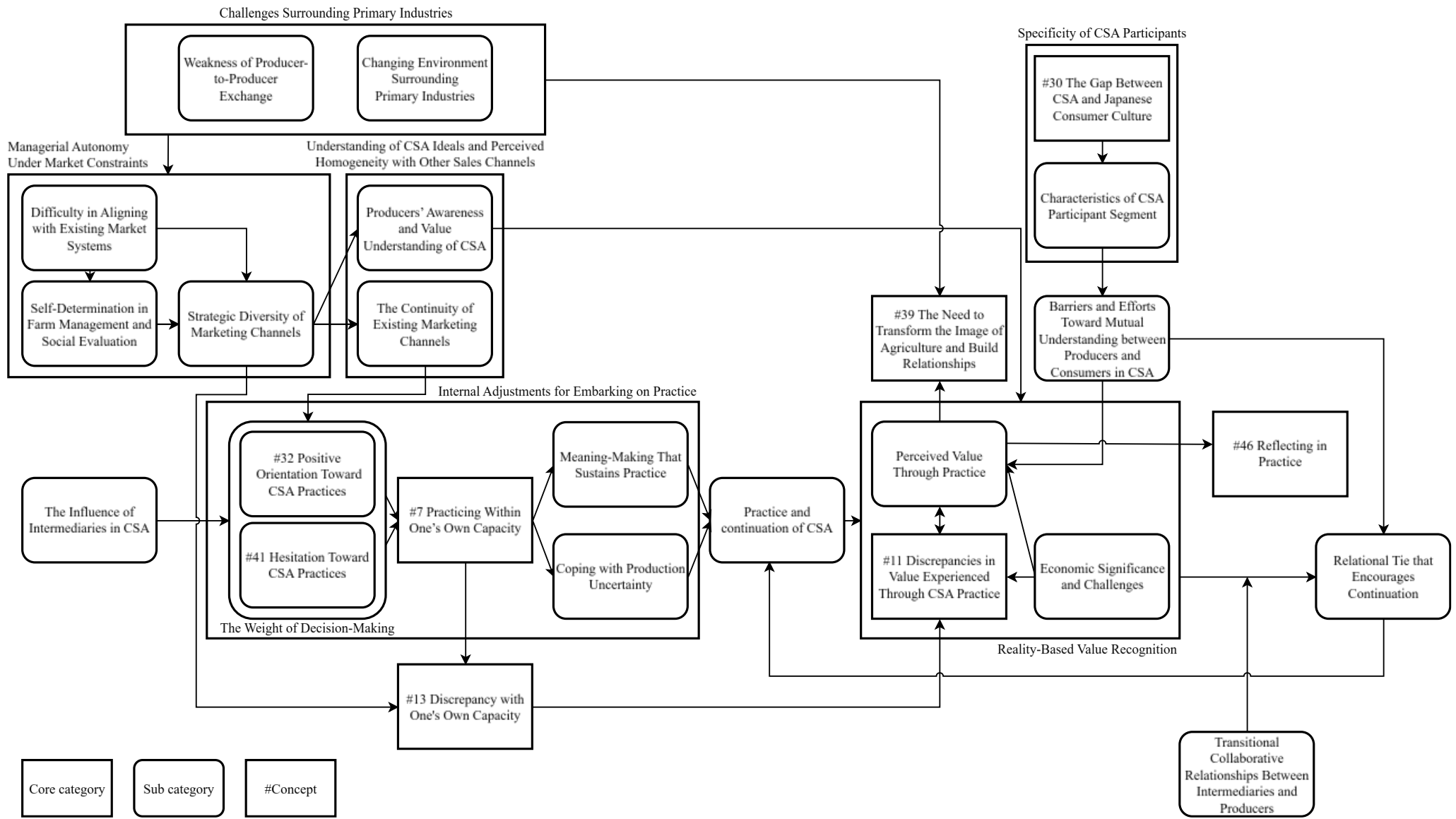

Fig. 1. Process of CSA practice by producers. The figure illustrates causal relationships, interrelations, and influences. Gray boxes represent core categories, rounded rectangles indicate subcategories, and rectangular boxes denote concepts.

To illustrate how producers' engagement in CSA unfolds as a dynamic process, the analytical results were summarized as a condensed storyline. Based on the analytical worksheets, the relationships among concepts, subcategories, and categories were systematically reviewed and integrated into a process diagram, from which a condensed storyline was developed. For readability, this section presents a concise version that captures the central mechanisms identified through the analysis. The full set of concepts and categories, along with the complete storyline, is available in Appendix B1.

Storyline:

Producers' decisions and actions related to CSA were influenced by intermediaries such as platform operators who provided organizational and logistical support, making participation feasible for those lacking direct access to consumers. Their involvement helped producers overcome hesitation and reinterpret CSA in ways consistent with their capacity and values. Through this process, producers developed practical adjustments, such as modifying delivery quantities or adding processed products, to cope with production uncertainty and align CSA practices with their own circumstances.

At the same time, producers faced structural tensions in balancing their self-determination in farm management with the constraints of existing market systems. Many sought to maintain autonomy by diversifying marketing channels and experimenting with management approaches less dependent on conventional distribution structures. These experiences shaped their understanding of CSA as both an alternative to, and partly overlapping with, existing food systems.

Once engaged in CSA, producers began to recognize new forms of value extending beyond economic returns to include social, educational, and relational aspects. While they regarded CSA participants as a particularly conscious consumer group, limited communication and differing expectations sometimes created barriers to mutual understanding. Nevertheless, these exchanges deepened producers' appreciation of CSA as a collaborative space for learning and connection.

Producers also acknowledged broader challenges in the primary industry, such as aging farmers, farmland abandonment, and shifts in agricultural policy, which reinforced CSA's role as a venue for sharing these realities with consumers. Favorable relationships with committed consumers and local partners, including restaurants and cafés, fostered strong relational ties that motivated producers to continue their CSA involvement. However, changes in intermediary structures occasionally disrupted trust and stability, reminding producers of the delicate balance between external support and managerial independence.

These findings reveal that the process of CSA implementation among producers unfolds not as a straightforward sequence but as an interrelated process supported and shaped by multiple actors at different stages. Producers' engagement evolves through a series of interconnected phases, from navigating relationships with existing markets and developing multiple marketing channels, to recognizing CSA and forming motivations for participation, and finally to the ongoing processes of practice and continuation.

Among these actors, intermediaries played a particularly critical role by providing entry points for CSA practice, accompanying producers throughout the implementation process, and facilitating consumer participation through outreach and coordination efforts. These intermediaries thus shaped both the initiation and continuity of CSA, forming part of the four key roles later integrated in the overall model of CSA community formation (Section 3.3). Furthermore, some producers disengaged from CSA when relational or managerial challenges outweighed the perceived benefits. Together, these insights provide the empirical foundation for understanding how producer-side dynamics contribute to the phased and actor-structured model of CSA community formation presented in Section 3.3.

3.2. CSA participation process for consumers

This section presents the results of the M-GTA analysis of the interview records with consumers (Fig. 2). The analysis generated 68 initial concepts. These concepts were compared and grouped into 22 subcategories (Table A3), each representing a higher level of abstraction that synthesizes multiple related concepts. Subsequently, the subcategories, together with unclassified concepts, were further integrated and abstracted to yield 9 overarching categories (Table A4). This hierarchical structuring, moving from concepts to subcategories and then to categories, provides an analytical framework for systematically understanding the factors that influence consumers' CSA participation.

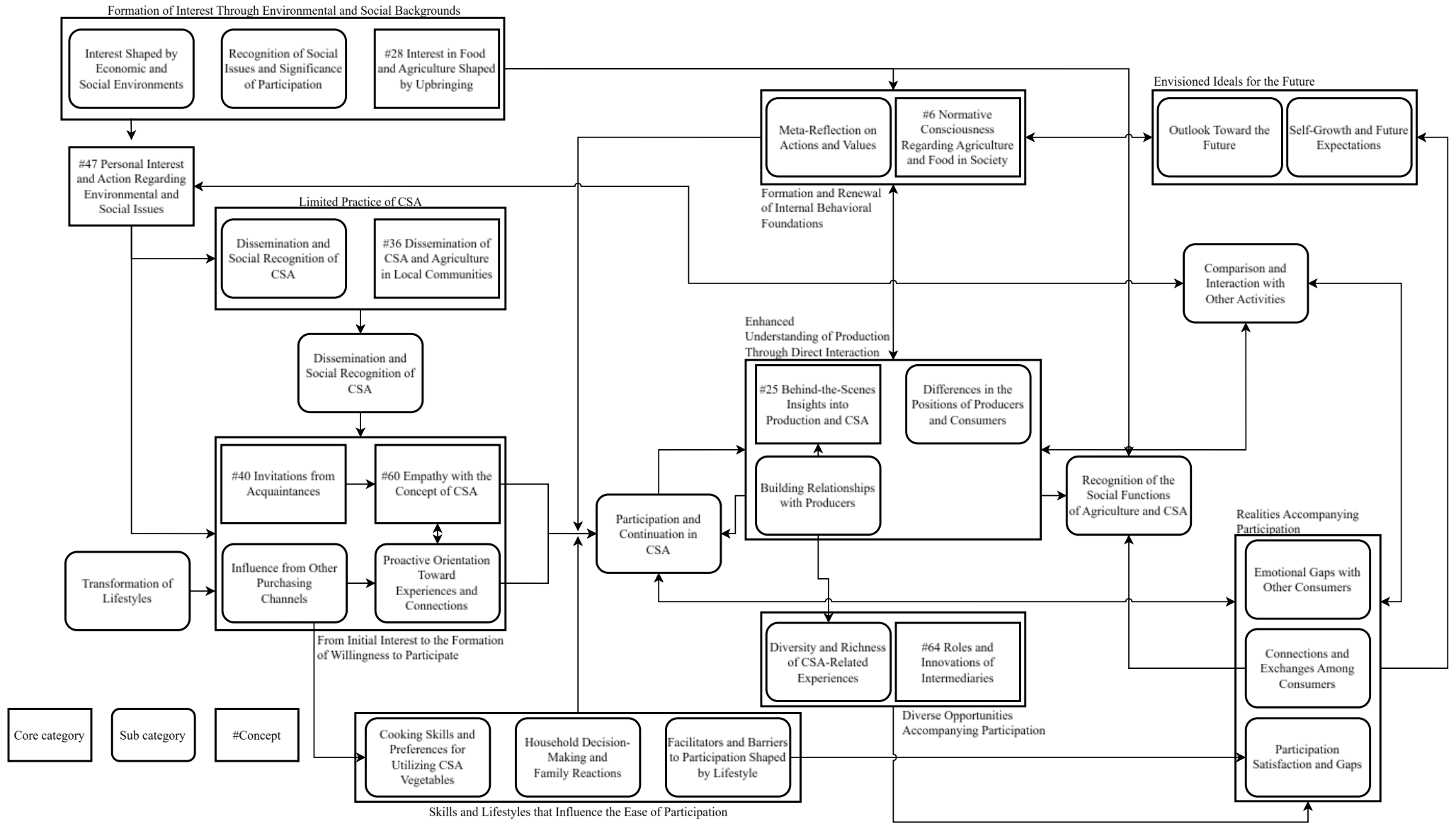

Fig. 2. Process of CSA participation by consumers. The figure illustrates causal relationships, interrelations, and influences. Gray boxes represent core categories, rounded rectangles indicate subcategories, and rectangular boxes denote concepts.

To illustrate how consumers' engagement in CSA unfolds as a dynamic process, the analytical results were summarized as a condensed storyline. Based on the analytical worksheets, the relationships among concepts, subcategories, and categories were systematically reviewed and integrated into a process diagram, from which a condensed storyline was developed. For readability, this section presents a concise version that captures the central mechanisms identified through the analysis. The full set of concepts and categories, along with the complete storyline, is available in Appendix B2.

Storyline:

Consumers' engagement with CSA begins with growing awareness shaped by their upbringing and broader social and environmental concerns. They encounter CSA through various information channels, empathy with its ideals, and invitations from acquaintances. Experiences with other purchasing systems, such as vegetable subscriptions, and a proactive interest in farming and community life further foster willingness to participate. These motivations are often reinforced by life transitions, including child-rearing, relocation, or the lifestyle changes brought by the COVID-19 pandemic.

Participation in CSA is influenced by consumers' skills and lifestyles, such as cooking ability, time availability, and previous experience with local food practices, which determine the ease of participation. Through farm visits and direct exchanges, they gain a deeper understanding of agricultural production, develop gratitude and respect toward producers, and recognize the differences in their respective positions.

As consumers become involved, they experience diverse opportunities created by intermediaries and producers, including farm events, food education, and community gatherings. These experiences enhance both satisfaction and connection among participants, while at times revealing emotional gaps or practical burdens. Through such experiences, CSA comes to be understood not only as a means of food procurement but also as a space for learning, interaction, and mutual support.

These encounters encourage consumers to compare CSA with other community or sustainability activities, fostering recognition of the broader social roles of agriculture and of CSA itself. They also lead to self-growth and a sense of future orientation. Reflecting on their own values and actions, consumers develop internal behavioral foundations that reinforce their willingness to continue participating. Positive relationships with producers and fellow members further encourage ongoing engagement and the deepening of commitment to CSA.

These findings show that consumers' engagement in CSA is shaped not only by their individual awareness and experiences but also by interactions with other actors throughout the process. Consumers' engagement unfolds through interrelated stages, beginning with growing awareness of environmental and social issues, as well as lifestyle transitions such as child-rearing, relocation, or shifts in work and living patterns that prompt greater attention to food and farming, followed by active information-seeking and recognition of CSA, the formation of willingness to participate, and finally the processes of participation and continuation. In addition to intermediaries who provide opportunities for learning and connection, producers' outreach and visibility before participation, such as farm events and direct marketing activities, often serve as key entry points that not only shape consumers' initial recognition of CSA but also lay the groundwork for building relationships after participation. As participation progresses, some consumers become more active by inviting acquaintances or helping with CSA-related activities, thereby supporting the continuity of engagement within the community. Others remain in more peripheral forms of participation with limited involvement, and some eventually disengage when changes in personal circumstances occur. Together, these differentiated stages, roles, and trajectories provide empirical insights into how consumer participation develops and stabilizes within CSA communities, forming the analytical foundation for both the identification of the four key actors and the phased structure of community formation presented in Section 3.3.

3.3. Explaining the Process of CSA Community Formation through Producer and Consumer Integration

In this study, we conceptualize "CSA community formation" as a process through which interactions around CSA evolve into a relationally embedded collective: relational, social, and cognitive proximity progressively deepen among producers, consumers, and intermediaries, and CSA is reconfigured from a one-off purchasing arrangement into an ongoing, shared practice. This study explains how such CSA communities are gradually formed through the integration of producer practices and consumer participation into a phased model (Fig. 3). The model illustrates how distinct key actors emerge at different stages of this process and how their interactions contribute to the progressive strengthening of relational, social, and cognitive proximity within CSA communities.

The construction of this model was grounded in the results of the M-GTA and complemented by iterative revisiting of interview transcripts and analytical worksheets to ensure close alignment with the data. We began by placing the producer and consumer process diagrams (Fig. 1 and Fig. 2) in parallel, organizing their trajectories by distinguishing the stages before and after participation and practice. Subsequently, interview transcripts and analytical worksheets were revisited iteratively to verify the consistency of interpretation and to refine the relationships among phases and actors. During this process, particular attention was given not only to identifying key actors repeatedly mentioned across interviews but also to conceptualizing both the stages of participation and practice and the roles these actors played within each emerging phase. The integration was further strengthened through iterative cycles of cross-checking and collaborative discussions among co-authors, ensuring both the interpretive validity and the stability of the final phased representation.

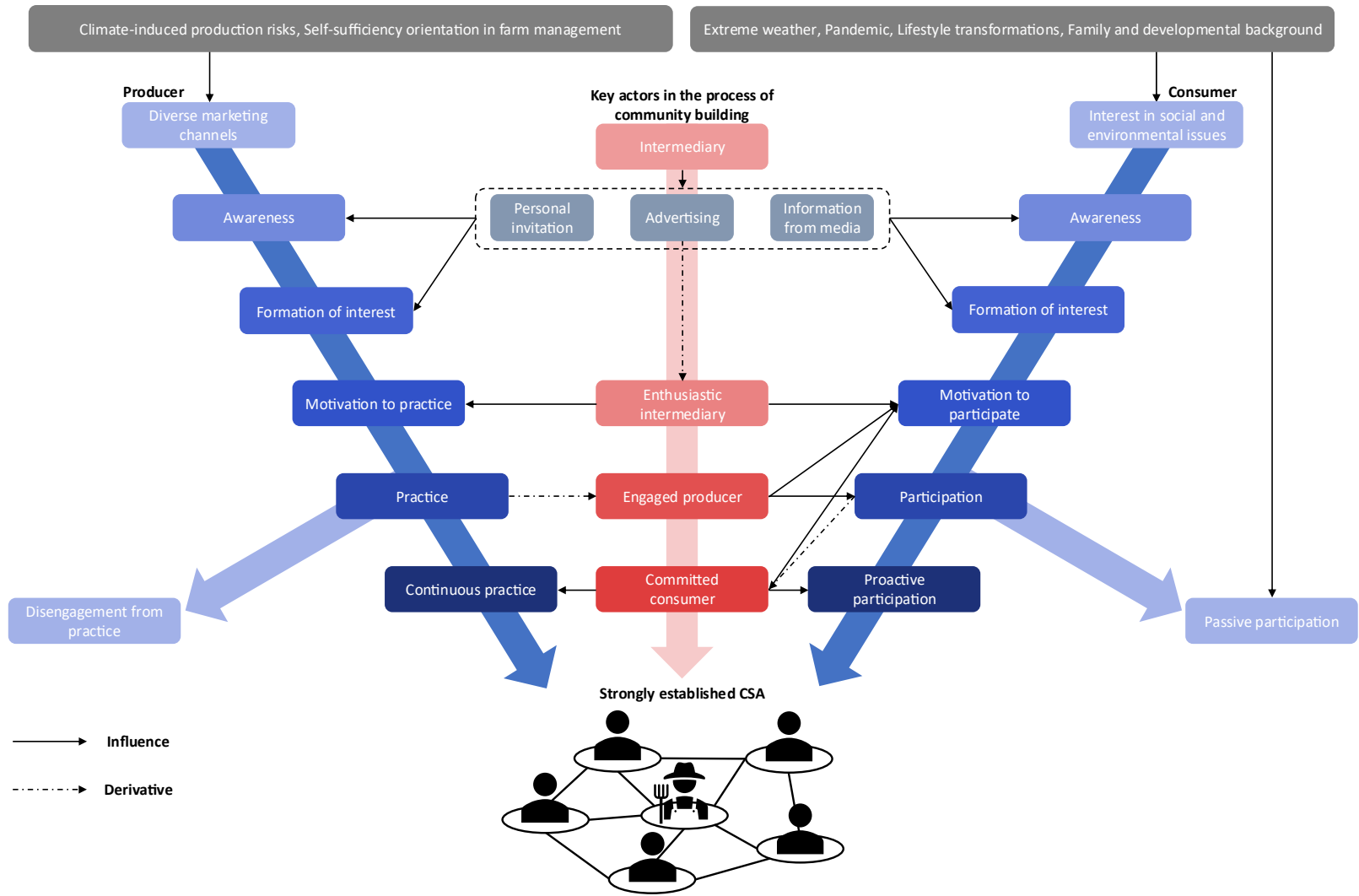

Fig. 3. Process of CSA community formation: integration of producer practices and consumer participation

3.3.1. A Six-Phase Process for Deepening the Practice of CSA

The six-phase model presented in Section 3.3 was developed by integrating the analytical results of producers (Section 3.1) and consumers (Section 3.2).

The producer-side analysis revealed processes related to the implementation and continuation of CSA practices, illustrating how engagement emerges from the pursuit of autonomy within existing market conditions and develops through adaptive decision-making and practice. In contrast, the consumer-side analysis elucidated the processes of awareness, interest formation, and motivation that lead to participation, as well as the evolving understanding and commitment that support continued involvement.

By synthesizing these complementary trajectories, the model conceptualizes how CSA communities evolve through six interrelated phases, ranging from the dispersed formation of interest and awareness to the consolidation of co-creative practice. These phases provide an empirically grounded framework that captures the gradual deepening of engagement within CSA. The following outlines each phase by presenting its name and conceptual definition.

**Phase 1: Dispersed Interest Formation**

In this initial phase, producers begin to show interest in diversifying distribution channels as a response to risks caused by climate change and market instability. Consumers, in turn, develop an interest in environmental and social issues, often triggered by life events such as child-rearing or by broader societal contexts, including news and public discourse on environmental and social challenges. At this stage, CSA is not yet recognized as a specific option by either group.

**Phase 2: Awareness**

In this phase, producers and consumers become aware of the concept of CSA for the first time. This recognition is often prompted by online content, social media, media coverage, or promotional activities by platform organizations functioning as intermediaries. At this point, awareness is still external and conceptual, rather than personalized.

**Phase 3: Formation of Interest**

At this stage, awareness is translated into personal relevance. Producers evaluate CSA in relation to their management identity and market diversification needs, while consumers connect CSA to their lifestyle shifts or values. In this stage, CSA is reframed from an abstract system into a practice aligned with one's own concerns.

**Phase 4: Motivation to Practice and Participation**

In this phase, motivation to practice and participate begins to take shape. While some individuals move directly from interest to action, many require additional stimuli, such as encouragement from enthusiastic intermediaries, stories from CSA producers, or invitations from existing members, to internalize CSA as something worth doing. The translation of CSA's values into personal significance plays a key role here.

**Phase 5: Practice and Participation**

This phase marks the beginning of concrete engagement with CSA. Participants begin to receive produce, join farm-based events, or engage in exploratory practice. Producers, for their part, advance their practices in an exploratory manner, drawing on their own knowledge and experience to adjust production and organizational approaches. These interactions promote direct encounters between producers and participants, facilitating mutual understanding and trust.

**Phase 6: Continuous Practice and Co-creative Participation**

Over time, some participants become highly engaged. These individuals take on roles beyond that of a recipient, co-organizing events, inviting new participants, or contributing to CSA governance and design. Their participation is rooted in shared values and relational commitment, transforming CSA into a co-created social institution rather than a service model. Through these practices, and by being influenced by the people they encounter along the way, participants build enduring relationships that sustain their continued involvement.

Through these six phases, a CSA characterized by strong social proximity is formed, fostering relationships between producers and consumers, as well as among consumers themselves. At the same time, varying degrees of involvement are observed within these phases. Some consumers remain in forms of "Passive Participation," maintaining nominal engagement with limited interaction, while some producers experience "Disengagement from Practice" due to managerial challenges. These variations illustrate that CSA participation is fluid and continually negotiated, rather than fixed or uniform.

3.3.2. The Role of Key Actors in Enabling CSA Participation

The identification of key actors was grounded in the analytical results of producers (Section 3.1) and consumers (Section 3.2).

In the producer analysis, intermediaries were found to play a pivotal role in initiating and sustaining CSA practices by providing organizational infrastructure and reducing entry barriers for producers. Meanwhile, the consumer analysis revealed that participation and continuation were supported by various relational influences, including producers' outreach and the involvement of active participants who encouraged others to join or stay engaged.

By synthesizing these insights, four types of actors were identified as particularly instrumental in enabling and sustaining CSA participation and, in turn, in driving the gradual consolidation of CSA communities as relationally embedded collectives: intermediaries, enthusiastic intermediaries, engaged producers, and committed consumers.

Each of these actors contributes to shaping and reinforcing relationships among producers, consumers, and the broader CSA network. The following subsections describe their distinctive characteristics and functions, drawing on interview excerpts to illustrate how they facilitate the processes of participation and practice.

**Intermediary:** The intermediary functions as an infrastructural actor that provides opportunities for CSA participation and practice to society. By utilizing websites, brochures, information sessions, social media, and mass media, they create an "entry point" to CSA and connect producers with consumers.

*"In that sense, I had been doing things on my own until now, but this time XXX (name of intermediary) organized an initiative and I was able to participate through it. Having that kind of support makes participation so much easier. I don't have to do the sales myself." (Producer B)*

*"When it came to actually organizing a CSA, we thought honestly that handling all the data collection and coordination would have been impossible while also farming and doing shipping work. But if another party provides the system side, then all we have to do is bring the vegetables, and that works for us." (Producer G, G')*

*"It just happened to come up on Instagram and, well, how should I put it… it was exactly the kind of thing I had always wanted to do deep down. I thought, wow, I didn't know such a system even existed (about CSA). It really struck a chord with me. I am not usually the type to jump at these things. (omission) Since it was so impressive, I decided to join from the very first session." (Consumer H)*

In this way, the intermediary played an essential role in lowering the entry barriers for producers by gathering a sufficient number of consumers and establishing CSA as a functional system. This relieved producers of the need to independently develop sales channels or manage membership and administrative tasks, thereby contributing significantly to the practice of CSA. At the same time, intermediaries also influenced consumers' awareness and interest by making CSA visible and accessible, helping latent concerns and values to be translated into actual participation.

**Enthusiastic intermediary:** Within the broader category of intermediaries, enthusiastic intermediaries emerge as actors who go beyond simple information provision. They play a critical role in translating the values and practices of CSA into the language of the stakeholders involved, making them more tangible and accessible. Working closely with farmers, they transform CSA's narratives, philosophies, and everyday practices into forms that can be shared with potential participants, thereby creating meaningful points of connection.

*"This year is our third year, so it must have been three years ago. At that time, there was a staff member, YYY (the enthusiastic intermediary), and since he was just starting up, he had incredible enthusiasm." (Producer G, G')*

*"Our customers, you know. Even our new customers were brought in by YYY, who explained everything. In the end, having someone who actually finds customers for us is, well, the biggest benefit for us." (Producer H, H')*

*"But YYY also came by regularly, checked in on us, and provided all kinds of support. This YYY really gave us a lot of follow up." (Producer H, H')*

In this way, enthusiastic intermediaries not only supported farmers by reducing the burden of recruiting new members but also sustained ongoing engagement through continuous follow-up and visible commitment. Their active involvement helped transform CSA from an abstract concept into a lived experience, thereby fostering trust, broadening participation, and strengthening the continuity of CSA practices.

**Engaged producer:** Beyond the practice of CSA, these actors engage in diverse activities such as farm events, educational initiatives, and community collaborations, all grounded in clear management policies and social ideals. These efforts provide opportunities for consumers to recognize the presence of farmers even before their direct involvement in CSA, thereby contributing to the formation of initial interest in participation.

*"Well, the reason I've continued for three years is because there's a genuine personal relationship. It's about having a proper connection, person to person. I also know both the importance and the difficulty of continuation, so part of it is that I am trying to take on that challenge." (Producer B)*

*"By that time, I already had a connection with Farmer C, through volunteer farm work. So, there was already a route in place. It was the combination of being able to pick up vegetables from a farmer I already knew and a restaurant known for its delicious dishes. Plus, the pick-up location was easy to get to from my home, so I thought, well, that could work for me, and I started participating from the second year." (Consumer D)*

*"In the case of Producer A, I think what has sustained my involvement is the personal qualities and original intentions behind their work. They also run a bookstore, and I felt they were really impressive people. That connection has been one of the reasons I have been able to*

*continue." (Consumer G)*

Taken together, these examples highlight how engaged producers act as relational anchors who embody values beyond agriculture itself. Their diverse activities and personal commitments generate recognition and trust that precede CSA participation, and this relational groundwork functions as an essential mechanism for transforming casual encounters into long-term involvement.

**Committed consumer:** Sharing the ideals of CSA and building on positive relationships with producers, they act not merely as consumers but as active agents. They engage in CSA in various ways, such as inviting acquaintances as new participants, actively joining farm events, and expressing opinions about the management of the CSA system. Such actions play a crucial role in developing CSA not only as a purchasing framework but also as a community.

*"Some acquaintances of Consumer E joined as new members. Consumer E was also part of another community called ZZZ, which is oriented toward activities related to a sustainable society and social good. About two or three people from that community joined, and in the second year, I would say the 'purity' of the members increased." (Producer C)*

*"Well, at the very beginning… let me see… yes, it was through an introduction from Consumer E that I joined. I had heard that Consumer E was participating in a platform, which is operated by XXX." (Consumer D)*

*"This year, one of my friends, who lives alone, said it would be nice if there were smaller quantities. So, I mentioned this to Producer C, and he created a single-person set." (Consumer E)*

These cases demonstrate how certain consumers behave as committed consumers, functioning as hubs that connect other consumers, external communities, and producers. As a result, CSA is strengthened beyond individual transactional relationships, becoming a more broadly interconnected social network.

4. Discussion
4.1. CSA Participation as Sensemaking and Relational Embedding

The findings show that both producers and consumers engage in a gradual process of making CSA meaningful and actionable under conditions of uncertainty. For producers, CSA practice begins with a series of internal adjustments, which are reflective assessments of feasibility, identity, and practical implications ([Internal Adjustments for Embarking on Practice]). These adjustments include negotiating psychological dilemmas, accepting production uncertainties, and exploring adaptive approaches suited to their farm context. Such dynamics can be interpreted through sensemaking theory (Weick, 1995), which explains how individuals translate ambiguous situations into concrete actions by drawing on their identities and lived contexts (Weick et al., 2005). In the nascent CSA settings in Japan, where outcomes and future continuity remain uncertain, farmers reduced uncertainty by constructing meanings rooted in everyday routines and self-understandings.

This sensemaking process also shaped how producers came to view CSA as an opportunity to improve the public image of agriculture. This observation aligns with previous studies showing that CSA allows producers to embody values and identities that extend beyond market considerations (Worden, 2004; Samoggia et al., 2019) and that CSA provides a setting capable of influencing consumer consciousness (Glinska-Newes et al., 2025).

Once such relationships began to take shape, producers further recognized relational gravity that foster the continuation (<Relational Ties that Encourage Continuation>). This relational dynamic corresponds to bonding social capital (Putnam, 2000). For example, one producer described how consumer interactions generated a sense of reciprocal commitment:

*"Our customers, well, it feels like we can't quit now. What made me happy this week was when they introduced things like, "Producer H and H' sent us these vegetables!" or "I made this dish!" Seeing that, the customers start chatting among themselves with comments like "That looks delicious!" and then asking things like "Can you teach me how to make it next time?" It's like a circle forming there, and I feel like I can't escape this" (Producer H and H').*

Consumers also engaged in a parallel process of making CSA meaningful in their own lives. Life-course triggers, including child-rearing, growing awareness of environmental concerns, and experiences during the pandemic, made CSA personally relevant at specific moments. As one participant noted,

*"I really didn't like the fact that I couldn't even grow a single vegetable. There are so many things I couldn't do, but at the very least I want to become someone who can grow vegetables in the future" (Consumer J).*

Although ethical and environmental motivations have been highlighted in previous studies (Brehm and Eisenhauer, 2008; Pole and Gray, 2013) ,our findings extend this understanding by demonstrating that life transitions and contextual shifts transform CSA from an abstract ideal into a personally meaningful and feasible option.

Everyday competencies, such as cooking skills and time availability, also influenced both participation and continuation. This point has received relatively little attention in earlier CSA studies Familiarity with subscription-based services appears to reduce typical uncertainties associated with CSA participation, which have been identified in earlier studies (Galt et al., 2019; de Souza, 2020).

Taken together, these findings indicate that CSA participation for both producers and consumers cannot be explained solely by normative motivations or economic rationality. Rather, engagement unfolds through a combined process of sensemaking and relational embedding. Individuals interpret CSA through their identities, routines, and life transitions, and these interpretations are reinforced through social proximity and repeated interaction. As relationships deepen, they strengthen motivation and stabilize continued participation, suggesting a dynamic feedback loop between meaning, practice, and relational ties.

4.2. Toward a Six-Stage Model of CSA Formation: Integrating Producers and Consumers

This section examines how two types of actors identified in the six stage process, intermediaries and committed consumers, shape the formation and continuity of CSA. While intermediaries function as gatekeepers that create entry points and reduce initial barriers, committed consumers strengthen cohesion and sustain participation over time. In the following subsections, we focus particularly on intermediaries, who provide entry points into participation, and committed consumer, who consolidates the community and drive its long-term development.

4.2.1. Intermediaries as Gatekeepers and Their Vulnerabilities

Intermediaries play a central role in SFSCs and sustainability transitions by building trust among stakeholders, coordinating actors, and supporting network formation (Kivimaa et al., 2019; Massuga, 2022; Sundqvist and Tuominen, 2024). The present case, however, demonstrates that when intermediary functions fail, especially due to disruptions in the platform or information system, communication becomes fragmented and dissatisfaction increases among both producers and consumers. In such situations, coordination may break down and lead to "disintermediation," a process in which producers and consumers bypass intermediaries and rebuild CSA on their own.

*"It had been going really well, but then suddenly the app stopped working, and after that everyone became fragmented. Some people contacted me via Instagram messages, others via LINE if we had exchanged contacts, and still others by email. As a result, I would sometimes miss things, and that was extremely stressful for me." (Producers H, H')*

*"We often said that the intermediary's designated service was hard to use. I even mentioned, why spend money on something like that, when you could just use tools that are already*

*available? We talked about it not only with the intermediary but also among members, and it came up all the time." (Consumer E)*

These findings are consistent with research indicating that intermediaries must secure legitimacy and adequate resources to operate effectively (Kivimaa et al., 2019; Sundqvist and Tuominen, 2024). Studies on innovation systems also highlight several structural vulnerabilities that can destabilize intermediary roles, including the neutrality paradox, functional ambiguity, and the funding paradox (Klerkx and Leeuwis, 2009).

In the present case, functional ambiguity was particularly evident. Consumers questioned the intermediary's legitimacy and perceived little added value in the designated platform because widely used tools such as LINE and Instagram were seen as equally sufficient. This perception weakened the intermediary's distinct role as a bridging actor. Notably, the intermediary-managed platform was discontinued at the end of the fiscal year, and responsibility for CSA coordination shifted to individual producers from April 2025 onward. This transition illustrates how intermediary withdrawal can occur in practice and how it increases the operational burden on producers while heightening the risk of disintermediation.

Such dysfunctions should not be interpreted as isolated operational failures. They reflect deeper structural weaknesses that erode intermediary legitimacy. Positioning the intermediary within the broader literature on innovation brokers makes these vulnerabilities clearer. As innovation brokers, intermediaries typically take on three functions identified by earlier studies (Klerkx and Leeuwis, 2009).

(1) Translating visions and articulating demands
(2) Balancing participants' burdens and managing innovation processes
(3) Orchestrating relationships and building collaborative networks

Strengthening these functions may involve diversifying funding sources to maintain neutrality, clarifying mandates to reduce ambiguity, and combining public support with member-based contributions to ensure continuity. These measures can enhance intermediary resilience and highlight that the formation and stability of CSA communities depend not only on producers and consumers but also on the reliability of gatekeeping actors.

At the same time, intermediaries mainly facilitate initial entry into CSA. Long-term consolidation and cohesion increasingly rely on actors who reinforce relational ties from within the community. This point underscores the importance of committed consumers, which is discussed in the next subsection.

### 4.2.2. Committed Consumers as Prosumers and Co-Creators

While intermediaries create entry points into CSA, committed consumers often become the central actors who sustain and strengthen the community. These participants move beyond passive receipt of food and take active roles in coordination, event organization, and knowledge-sharing. Such involvement aligns with the concept of prosumers, who shape their own consumption experiences through value-creating practices and participate in both production and consumption (Toffler, 1980; Xie et al., 2008; Ritzer and Jurgenson, 2010).

Prior studies indicate that active engagement with producers and other members encourages prosumption and fosters cooperative forms of participation within CSA (Espelt, 2020). Opportunities that extend beyond food acquisition, such as farm work or community activities, also support a shift from passive participation to co-creation (Carolan, 2017; Galt et al., 2019). Identification with farmers' values further motivates members to deepen their involvement (Brehm and Eisenhauer, 2008; Russell and Zepeda, 2008).

This study advances these discussions in two ways. First, it empirically shows that prosumers emerge through the community-formation process itself. Rather than beginning as pre-existing or individually motivated actors, CSA participants become prosumers gradually through repeated interaction and embodied participation. They initially join as ordinary consumers and later take on coordinating or mediating roles. Their increasing involvement reflects experiential learning and the internalization of shared values, developed through repeated interactions and shared practices.

Second, the study highlights that prosumers function as agents of community cohesion. These participants do not simply sustain their own engagement but actively strengthen the social fabric of the CSA by recruiting like-minded members and fostering trust. As one producer observed, *"purity of the members increased. (Producer C)"* when committed consumers invited others who shared similar orientations. This illustrates how prosumers deepen bonding social capital (Putnam, 2000), the shared trust, reciprocity, and norms that reinforce internal solidarity. Conceptually, such community "refinement" can be understood as value alignment and boundary reinforcement (Lamont and Molnár, 2002) through which the group reaffirms what it collectively stands for.

In this sense, prosumers in CSA serve not only as co-creators of functional or economic value but also as stewards of relational and moral order, cultivating the trust and shared meanings that sustain the collective over time. This perspective extends prior understandings of prosumption from an individual-level dual role to a community-embedded process that reinforces the continuity and cohesion of CSA as a living social institution. It also complements the gatekeeping functions of intermediaries discussed in the previous subsection by showing how internal relational work supports long-term community stability.

### 4.3. Positioning the Six-Stage CSA Model within Community-Formation Theories

The six-phase CSA model developed in this study integrates producer and consumer trajectories to illustrate the progression from dispersed interest to co-creative continuation. When compared with existing theories of community formation, both its uniqueness and points of convergence become clear. Classic group development theory conceptualizes community evolution through the stages of forming, storming, norming, and performing (Tuckman, 1965). The latter phases of our model, "Practice and Participation" and "Continuous Practice and Co-creative Participation," parallel the norming and performing stages, highlighting processes through which participants deepen their roles and strengthen collaboration.

In contrast to classical models, the CSA model explicitly identifies pre-phases such as "Dispersed Interest Formation," "Awareness," "Formation of Interest," and "Motivation to Practice and Participation." These stages illustrate how individuals who are not yet members of any collective gradually converge around shared concerns and begin shaping a CSA community. In this sense, the model complements and extends existing community-formation theory by incorporating the antecedent processes through which dispersed actors become a collective.

Furthermore, the processes of "Disengagement from Practice" and "Withdrawal" observed in this study resonate with the "Adjourning" phase proposed later in group development research (Tuckman and Jensen, 1977). This correspondence indicates that the six-phase model captures not only the formation and consolidation of CSA communities but also the dynamics of disengagement and dissolution.

At the same time, the findings of this study are limited to the internal dynamics of single communities. Prior research on community resilience has emphasized that networks across multiple communities can function as safety nets, facilitating the exchange of practices and experiences while compensating for individual vulnerabilities (Jia et al., 2024). Future research is therefore needed to examine how inter-community connections may enhance the resilience of CSA, and how such linkages can be effectively designed and sustained.

### 5. Conclusion

This study theorized the trajectories of both producers and consumers in CSA in Japan, where practices remain limited and fragmented, and demonstrated how their interactions collectively shape community formation. Producers were shown to navigate uncertainty through internal adjustments and acts of sensemaking that provided coherence in contexts where established knowledge of CSA was scarce. Consumers' engagement was shaped not only by normative concerns such as environmental and social values but also by the practical feasibility of participation and turning points in their life courses. By integrating these trajectories, we proposed a six-phase model of CSA development that traces the progression from dispersed interest formation, awareness, and interest formation, to motivation, practice, and co-creative continuation. The model highlights the pivotal roles of intermediaries, enthusiastic intermediaries, engaged producers, and committed consumers in advancing CSA as more than a marketing channel and positioning it as a co-created social institution.

At the same time, this study has certain limitations. The qualitative design and relatively small sample size constrain the generalizability of the findings, while the Japanese context and reliance on a specific intermediary platform may limit their transferability. Nevertheless, through thick description and contextual detailing, this study sought to enhance the transparency and interpretability of the findings, thereby allowing readers to assess their applicability within their own contexts. Potential selection biases from recruitment methods, the cross-sectional nature of the data, and the positionality of the researchers also need to be considered.

Future research should not only validate the six-phase model across different contexts through longitudinal and comparative studies but also examine what types of ideal elements characterize people prior to joining CSA in regions where practices are still emergent. Such investigations would help to clarify how individuals with certain orientations, skills, or life situations are more likely to become participants and eventually committed members. By combining this line of inquiry with studies on intermediaries, relational capital, and the interplay between normative motivations and practical constraints, scholars and practitioners can design strategies that foster resilient, inclusive, and sustainable CSA communities even in areas where existing knowledge and practices are still developing.

6. Declaration of AI and AI-assisted technologies in the writing process

In writing this paper, after preparing a full text draft in the lead author's non-English native language, the authors used OpenAI's artificial intelligence language model, ChatGPT, to prepare an English draft. After using this tool, the authors carefully reviewed and edited the generated content to ensure the flow, logic, and accuracy of the text, making additions as necessary. Therefore, full responsibility for the content of the publication rests with the authors.

7. CRediT authorship contribution statement

**Sota Takagi:** Conceptualization, Methodology, Resources, Formal analysis, Investigation, Writing – Original Draft, Visualization. **Miki Saijo:** Validation, Writing - Review & Editing. **Takumi Ohashi:** Supervision, Project administration, Validation, Writing - Review & Editing, Funding acquisition. All authors contributed critically to the drafts and gave final approval for publication.

8. Declaration of competing interest

The authors declare that they have no competing interests.

9. Acknowledgments

This work was supported by JSPS KAKENHI Grant Number JP24K17976.

Table A1. Subcategories generated by concepts comparison. (Producer)

| # | Subcategory | Definition | Encapsulated concepts |
|---|---|---|---|
| 1 | Difficulty in Aligning with Existing Market Systems | Japan's existing market distribution system is structured on the premise of large-scale, single-crop production, making it poorly compatible with diversified cropping and organic farming. Consequently, producers perceive difficulties in ensuring the sustainability of their farm management when relying on market shipment and existing supply chains. | #2 Challenges in Adapting to the Existing Food Supply System<br>#6 Mismatch between Japan's Market Structure and Organic Farming |
| 2 | Strategic Diversity of Marketing Channels | Against the backdrop of diversified and organic farming, producers strategically select and utilize multiple sales channels according to their management styles. Opportunities for expanding sales outlets often arise when particular activities or the unique characteristics of producers gain external recognition. | #5 Balancing Diverse Marketing Channels<br>#9 Fostering Producers' Awareness of Opportunity Utilization<br>#27 Opportunities for Developing Sales Outlets |
| 3 | Self-Determination in Farm Management and Social Evaluation | Producers establish marketing channels and management policies based on their own principles, placing strong emphasis on these self-defined approaches in their farm management. At the same time, they often feel that such unique initiatives and philosophies are not sufficiently recognized or valued by society or consumers, creating a gap between their self-determined stance and external evaluation. | #19 Producer Stance in Farm Management<br>#29 The Gap Between Producers' Initiatives and Social Recognition/Evaluation |
| 4 | Producers' Awareness and Value Understanding of CSA | Producers' understanding and appreciation of CSA are shaped through diverse recognition pathways and personal experiences. Some producers are familiar with the concept and mechanisms of CSA even before engaging in its practice, showing interest in its normative values such as building connections with consumers. In contrast, other producers become aware of CSA only after participating, gradually recognizing its value through hands-on experience and continued practice. | #10 Producers' Perception of the Normative Values of CSA<br>#21 Diversity in Producers' Pathways to CSA Awareness |
| 5 | The Influence of Intermediaries in CSA | Intermediaries play a key role in providing producers with new marketing outlets and promotional opportunities, thereby reducing the burden of customer acquisition. In addition, direct solicitation and proactive engagement from intermediaries exert a strong influence on producers' participation in and continuation of CSA. | #4 Direct Solicitation from Intermediaries<br>#8 CSA Practice Opportunities Provided by Intermediaries<br>#44 The Presence of Enthusiastic Intermediaries |

| 6 | The Continuity of Existing Marketing Channels | Some producers perceive CSA as essentially no different from their existing sales channels, positioning it in a similar way to other marketing outlets. Such recognition facilitates the identification of commonalities between conventional sales methods and CSA, thereby lowering both the psychological and practical barriers to entering CSA as a new marketing channel. | #1 Perceived Homogeneity Between CSA and Other Sales Channels<br>#3 Ease of Participation Derived from Prior Sales Experience |
|---|---|---|---|
| 7 | The Weight of Decision-Making | While some producers demonstrate a positive willingness to engage in CSA practices, others experience hesitation or anxiety regarding their implementation. The weight of decision-making thus varies among producers, reflecting their differing levels of motivation and concern. | #32 Positive Orientation Toward CSA Practices<br>#41 Hesitation Toward CSA Practices |
| 8 | Meaning-Making That Sustains Practice | Producers strongly recognize the sense of responsibility for supply that arises from advance payment contracts, which serves as a motivating factor for enhancing their willingness to engage in CSA. In addition, external factors beyond CSA also function as motivations that support producers' practice and continuation of CSA. | #18 Meaning-Making That Drives Willingness for Practice and Continuation<br>#42 Supply Responsibility Stemming from Advance Payment Contracts and the Enhancement of Producers' Motivation |
| 9 | Coping with Production Uncertainty | Producers are acutely aware of the uncertainties in production that arise from seasonality and weather-related factors in vegetable cultivation. In particular, they often feel pressure to secure sufficient yields at the time of regular share distribution. To address these anxieties and difficulties, producers make flexible adjustments in the content and composition of the shares, thereby ensuring stable provision to consumers. | #22 Production Uncertainty and the Pressure of Distribution Timing<br>#31 Flexible Arrangements in Share Composition |
| 10 | Perceived Value Through Practice | Producers experience diverse forms of value through CSA practice, such as strengthened relationships with consumers and the expansion of marketing opportunities. | #23 Social Value Perceived Through CSA<br>#36 Potential Experienced During CSA Practice |
| 11 | Economic Significance and Challenges | Both producers and intermediaries recognize the benefits and stability of securing income through advance payment contracts as a tangible economic value. At the same time, they acknowledge the difficulty of generating sufficient profit from CSA | #16 Economic Challenges of CSA<br>#43 Monetary Value Perceived Through CSA |

| | | alone, which necessitates supplementing their income through other marketing channels. | |
|---|---|---|---|
| 12 | Transitional Collaborative Relationships Between Intermediaries and Producers | For both intermediaries and producers, CSA remains at a transitional stage in which relationships and operational conditions are not yet fully established. While producers are influenced by the circumstances and conditions set by intermediaries, they continue to seek more sustainable forms of collaboration. At the same time, certain operational conditions imposed by intermediaries act as disincentives to producers' willingness to continue, highlighting the ongoing need for adjustments toward stabilizing the collaborative relationship. | #14 Exploration and Adjustment Processes in Intermediary–Producer Collaboration #33 Intermediary Circumstances and Conditions Hindering the Continuation of CSA |
| 13 | Relational Ties that Encourage Continuation | The positive relationships developed between producers, stakeholders, and consumers through CSA activities function as a driving force that sustains producers' engagement in CSA practices. These relationships act as a form of "relational ties," enhancing producers' willingness to continue their involvement. | #15 Relational Tie Cultivated Through CSA Activities #25 Positive Relationships with Local Hubs |
| 14 | Barriers and Efforts Toward Mutual Understanding Between Producers and Consumers in CSA | In the practice of CSA, there are evident gaps in recognition and understanding of the system and its values between producers and consumers. Consumer awareness of CSA remains limited, and communication between producers and consumers occurs only within a narrow scope, leaving many barriers to achieving mutual understanding. At the same time, with some consumers, communication has enabled the sharing of values and knowledge, indicating ongoing efforts to deepen mutual understanding. | #26 Low Level of CSA Awareness Among Japanese Consumers #34 Consumer Understanding Fostered Through Communication #35 Barriers to Mutual Understanding with CSA Participants |
| 15 | Characteristics of CSA Participant Segment | The consumer segments participating in CSA tend to be skewed toward highly conscious segments, while securing target consumer groups has become increasingly difficult due to broader social trends and demographic changes. Moreover, participation pathways are diversifying, encompassing not only voluntary participation but also referrals from existing members and intermediaries. | #12 Highly Conscious Consumers #28 Diverse Participation Pathways of Consumers #45 Absence of Target Consumers |
| 16 | Changing Environment Surrounding | In the field of primary industries, challenges such as population aging, a shortage of | #38 Farm Exit and Labor Volatility |

| | Primary Industries | successors, and the volatility of labor stemming from dependence on foreign workers have become pressing concerns. At the same time, interest in primary industries is gradually increasing, particularly among younger generations. Moreover, government policies on regional revitalization and broader social trends toward sustainability serve as external factors that stimulate greater interest in agriculture and CSA. | #37 Growing Interest in Primary Industries #40 External Factors Enhancing Interest in Primary Industries |
|---|---|---|---|
| 17 | Weakness of Producer-to-Producer Exchange | Producers tend to remain confined to their own bases and marketing channels, with limited interaction with producers from other CSA sites or with producers in the same region. Furthermore, generational differences and variations in farming styles contribute to the weakness of local producer-to- producer exchange, resulting in relatively inactive information sharing and collaboration among producers. | #20 Limited Exchange with Local Producers #24 Connections with Producers from Other CSA Sites |
| 18 | Practice and continuation of CSA | Producers' CSA practices and continuity | #47 CSA in Practice #48 Continuing the Practice |

Table A2. Categories generated by subcategories and concepts comparison. (Producer)

| # | Category | Definition | Encapsulated subcategories and concepts |
|---|---|---|---|
| 1 | Managerial Autonomy Under Market Constraints | Producers recognize the challenges posed by Japan's market structure, which is oriented toward single-crop production and poorly suited for the distribution of small-scale, diversified, or organic farming products. In response, they adopt a proactive stance toward developing and securing their own sales channels. While strategically selecting CSA alongside other outlets as part of their marketing strategy and pursuing greater managerial autonomy, producers also face administrative burdens associated with developing new channels as well as insufficient social recognition and evaluation of their efforts. | Difficulty in Aligning with Existing Market Systems Self-Determination in Farm Management and Social Evaluation Strategic Diversity of Marketing Channels |
| 2 | Understanding of CSA Ideals and Perceived Homogeneity with Other Sales Channels | Although producers acknowledge the normative values of CSA, such as sustainability and contributions to local communities, some do not perceive CSA as | Producers' Awareness and Value Understanding of CSA |

| | | fundamentally different from conventional sales methods. Instead, they position CSA as an extension of their existing marketing channels. Pathways of CSA awareness and prior sales experiences help lower barriers to CSA entry, leading producers to adopt CSA practices without treating them as exceptional or distinctive. | The Continuity of Existing Marketing Channels |
|---|---|---|---|
| 3 | Internal Adjustments for Embarking on Practice | When engaging in CSA, producers undergo an internal process of adjustment in which they reflect on the significance of practice and their own capacity. This involves negotiating psychological dilemmas, accepting physical constraints and production uncertainties, and exploring feasible ways of implementation suited to their circumstances. | #7 Practicing Within One's Own Capacity<br>The Weight of Decision-Making<br>Meaning-Making That Sustains Practice<br>Coping with Production Uncertainty |
| 4 | Reality-Based Value Recognition | Through CSA practice, producers come to experience social values and enjoyment, such as strengthened relationships with consumers and the expansion of marketing opportunities. At the same time, they sometimes perceive discrepancies between the normative values of CSA and their actual experiences. Moreover, producers simultaneously recognize realistic challenges—such as the difficulty of achieving economic self-sufficiency through CSA and the financial responsibility associated with advance payment contracts—which coexist with motivations to continue CSA. | #11 Discrepancies in Value Experienced Through CSA Practice<br>Perceived Value Through Practice<br>Economic Significance and Challenges |
| 5 | Specificity of CSA Participants | CSA participants tend to be concentrated among socially conscious groups or individuals with particular value orientations, which diverges from the broader consumer culture and purchasing behaviors in Japanese society. Participation pathways are diversifying, encompassing not only voluntary interest but also referrals through acquaintances and intermediaries. Cultural backgrounds and differences in value orientations exert significant influence on the extent to which CSA can expand to a broader consumer base. | #30 The Gap Between CSA and Japanese Consumer Culture<br>Characteristics of CSA Participant Segment |
| 6 | Challenges Surrounding Primary Industries | The development of primary industries, including CSA practices, is entangled with both structural challenges, such as the lack of exchange among producers, and broader social and institutional issues, including population aging, a shortage of | Weakness of Producer-to-Producer Exchange<br>Changing Environment Surrounding Primary Industries |

| | | successors, and dependence on foreign labor. At the same time, the growing interest of younger generations in primary industries, along with policy and cultural trends surrounding regional revitalization, may provide entry points for addressing these challenges and exploring new possibilities. | |

Table A3. Subcategories generated by concepts comparison. (Consumer)

| # | Subcategory | Definition | Encapsulated concepts |
|---|---|---|---|
| 1 | Participation and Continuation in CSA | Refers to consumers' initial participation in CSA and their continued engagement over time. | #30 Participation in CSA<br>#68 Continuation of CSA Engagement |
| 2 | Interest Shaped by Economic and Social Environments | Economic and social contextual factors, such as rising prices and generational value orientations, affect consumers' considerations when deciding whether to participate in CSA. | #1 Acceptance of Prices<br>#16 Influence of Social and Generational Environments and Experiences |
| 3 | Influence from Other Purchasing Channels | Experiences with other purchasing channels, such as subscription-based purchases or direct sales from producers, shape consumers' interest in face-to-face interactions with producers and their willingness to participate in CSA. | #7 Access to Fresh Vegetables<br>#19 Degree of Interest in Organic or Additive-Free Products<br>#20 Shift Toward Subscription-Based Purchasing Styles<br>#49 Purchasing Food at Close Proximity to Producers |
| 4 | Outlook Toward the Future | A perspective in which participants consciously connect their CSA involvement and post-participation activities to future life and career choices. | #2 Life-Stage Transitions and Future Outlook<br>#18 Future Outlook Informed by CSA Experiences |
| 5 | Differences in the Positions of Producers and Consumers | Through exposure to producers' efforts and circumstances, participants become aware of the differences in roles and positions between themselves and producers. This psychological process leads them to develop feelings of gratitude and respect toward producers while also prompting reflection on their own stance toward | #8 Awareness of Differences and Commonalities with Producers<br>#9 Drawing Boundaries Around Participation<br>#15 Developing Gratitude and Respect Through |

| | | participation. | Recognition of Producers' Efforts |
|---|---|---|---|
| 6 | Cooking Skills and Preferences for Utilizing CSA Vegetables | Possessing the skills and preferences necessary to enjoy and make use of the diverse vegetables delivered through CSA shares. | #21 Cooking Skills and Preferences for Vegetables<br>#22 Opportunities to Obtain Cooking Information from Producers |
| 7 | Proactive Orientation Toward Experiences and Connections | Demonstrating active behaviors and exploratory intentions to seek experiences with producers and farming. | #3 Willingness for Direct Interaction Fostered Through Subscription Purchases<br>#4 Exploration of Opportunities for Farming Experiences<br>#46 Orientation Toward Building Connections with Producers<br>#39 Consciousness of Contributing to Producers and Agriculture |
| 8 | Meta-Reflection on Actions and Values | A reflective process in which participants objectively reconsider their own behaviors and choices, becoming aware of new values or behavioral patterns, and engaging in inner transformation. | #51 Changes in One's Own Behaviors and Thinking<br>#56 Meta-Awareness of One's Own Actions |
| 9 | Building Relationships with Producers | Relationships and trust are cultivated through consumers' recognition of producers' personalities and attentiveness, as well as their awareness of producers' activities even prior to participation. | #23 Touchpoints with Producers<br>#35 Consideration and Attentiveness of Producers<br>#45 Producers' and Staff Members' Personalities |
| 10 | Participation Satisfaction and Gaps | After joining CSA, the satisfaction derived from experiences, and the gaps between expectations before participation and actual experiences, affect consumers' willingness to continue. | #5 Subjective Value Perception Through Experience<br>#31 Knowledge and Learning Gained Through Agriculture and CSA Participation<br>#44 Discrepancies Between Values/Expectations and One's Own Behavior<br>#57 Differences from Pre-Participation |

| | | | Expectations |
|---|---|---|---|
| 11 | Self-Growth and Future Expectations | The struggles and adversities experienced in CSA become opportunities for self-growth, fostering an orientation toward learning and leading participants to hold expectations for the future direction of CSA. | #13 Expectations for the Future Development of CSA<br>#29 Awareness of Self-Growth Through Struggles and Adversities |
| 12 | Comparison and Interaction with Other Activities | Through comparisons and complementary relationships with other farming experiences and activities outside CSA, participants re-recognize the value of CSA and reshape their relationships with producers. | #41 Farming Experiences Outside CSA<br>#42 Comparisons with Non-CSA Agricultural Activities<br>#54 Synergies with Other Activities<br>#52 Recognition and Prior Contact with CSA Producers' Activities Before Participation |
| 13 | Diverse Information Channels | When considering CSA participation, consumers rely on various sources of information, including media exposure and points of contact with environmental or social issues, which act as external influences. | #43 Information Gathering About CSA<br>#48 Work-Related Contact with Environmental and Social Issues<br>#55 Influence from Media |
| 14 | Recognition of Social Issues and Significance of Participation | Participants' awareness of uniquely Japanese forms of environmental consciousness and concerns regarding the opacity of existing supply chains contributes to their recognition of the broader significance of CSA participation. | #61 Opacity of Existing Supply Chains<br>#63 Environmental Consciousness in Japan |
| 15 | Household Decision-Making and Family Reactions | Household decision-making structures, family members' understanding, and changes within the family influence participation in CSA. | #32 Triggers for Intra-Family Conversations About CSA<br>#37 Ease of Participation Shaped by Family Characteristics and Household Decision-Making Structures<br>#38 Family Changes and Reactions |
| 16 | Diversity and Richness of CSA-Related Experiences | The richness of CSA experiences is shaped by the diversity of share contents as well as additional opportunities for engagement provided by producers. | #10 Opportunities Beyond CSA Provided by Producers |

| | | | #24 Variety in CSA Shares |
|---|---|---|---|
| 17 | Emotional Gaps with Other Consumers | Emotional discomfort or differences in motivation compared with other consumers affect participants' experiences with CSA. | #12 Frustration Toward Consumers Whose Sensibilities Diverge from One's Own<br>#14 Differences in Motivation and Sensibilities with Other Consumers |
| 18 | Recognition of the Social Functions of Agriculture and CSA | Awareness of the broader social roles of agriculture, such as its non-business dimensions and the role of CSA in ensuring food security during emergencies. | #11 Non-Business Dimensions of Agriculture<br>#17 Recognition of Agriculture's Social Functions<br>#65 Food Security in Times of Crisis |
| 19 | Connections and Exchanges Among Consumers | Formation of networks and mutual empathy among consumers through information exchange and shared experiences. | #33 Consumer Networks and Empathy Formation<br>#34 Information Exchange Among Participants |
| 20 | Facilitators and Barriers to Participation Shaped by Lifestyle | Household environments, lifestyles, and residential locations influence the ease of CSA participation, creating both facilitating factors and barriers. | #26 Temporal and Geographic Constraints on CSA Participation<br>#27 Compatibility with One's Lifestyle<br>#53 Pragmatic Acceptance of the System<br>#58 Presence of Share-Splitting Partners |
| 21 | Dissemination and Social Recognition of CSA | Factors influencing CSA's dissemination, including its level of public recognition, the degree to which it is socially problematized, and reactions when introduced to others. | #50 Reactions When Introducing CSA to Acquaintances<br>#59 Public Awareness of CSA<br>#62 Current Challenges of CSA from an External Perspective |
| 22 | Transformation of Lifestyles | Changes in lifestyle patterns due to child-rearing, relocation, or events such as COVID-19 pandemic. | #66 Periods of Child-Rearing<br>#67 Changes in Living Environments |

Table A4. Categories generated by subcategories and concepts comparison. (Consumer)

| # | Category | Definition | Encapsulated subcategories and concepts |
|---|---|---|---|
| 1 | Envisioned Ideals for the Future | A perspective in which participation in agriculture is oriented toward the future, with participants constructing ideal images of their lives and careers based on self-growth and learning gained through CSA involvement. | Outlook Toward the Future<br>Self-Growth and Future Expectations |
| 2 | Realities Accompanying Participation | The realities recognized through CSA participation include satisfaction derived from the experience, gaps between expectations and actual outcomes, emotional distance from other consumers, and perceptions shaped through exchanges among participants. | Participation Satisfaction and Gaps<br>Emotional Gaps with Other Consumers<br>Connections and Exchanges Among Consumers |
| 3 | Formation of Interest Through Environmental and Social Backgrounds | The process by which interest in CSA and agriculture emerges from awareness of economic or social issues, as well as from personal backgrounds such as upbringing and early experiences with food and farming. | Interest Shaped by Economic and Social Environments<br>Recognition of Social Issues and Significance of Participation<br>#28 Interest in Food and Agriculture Shaped by Upbringing |
| 4 | Limited Practice of CSA | The influence of information-gathering pathways, recognition of the social significance of agriculture and CSA, and understanding of the current level of CSA dissemination on consumers' participation decisions. | Dissemination and Social Recognition of CSA<br>#36 Dissemination of CSA and Agriculture in Local Communities |
| 5 | Enhanced Understanding of Production Through Direct Interaction | A process in which participants deepen their understanding of production by directly engaging with producers, learning about their activities and efforts, and thereby building relationships that enhance the resolution of their perception of farming. | Differences in the Positions of Producers and Consumers<br>Building Relationships with Producers<br>#25 Behind-the-Scenes Insights into Production and CSA |
| 6 | Formation and Renewal of Internal Behavioral Foundations | An internal process in which participants cultivate normative consciousness regarding agriculture and food, while also reflecting on and revising their own actions and values, leading to a recognition of personal change. | Meta-Reflection on Actions and Values<br>#6 Normative Consciousness Regarding Agriculture and Food in Society |
| 7 | Diverse Opportunities Accompanying Participation | CSA activities provide participants with diverse experiences, as well as opportunities shaped by the program design and operational innovations introduced by intermediaries. | Diversity and Richness of CSA-Related Experiences<br>#64 Roles and Innovations of Intermediaries |
| 8 | From Initial Interest to the Formation of Willingness to Participate | Before deciding to join CSA, consumers' willingness to participate is cultivated through multiple early points of contact, including sales channels, experiential opportunities, and interpersonal invitations. | Influence from Other Purchasing Channels<br>Proactive Orientation Toward Experiences and Connections<br>#40 Invitations from Acquaintances<br>#60 Empathy with the Concept of CSA |
| 9 | Skills and Lifestyles that Influence the Ease of Participation | In CSA participation, participants' cooking skills and preferences, household decision-making structures, and temporal or geographic living conditions can function both as facilitators and as barriers to participation. | Cooking Skills and Preferences for Utilizing CSA Vegetables<br>Household Decision-Making and Family Reactions<br>Facilitators and Barriers to Participation Shaped by Lifestyle |

**Appendix B1 Storyline for producers**

In the following presentation, categories are indicated as [Category], subcategories as <Subcategory>, and concepts as *concept*.

The decision-making of producers regarding the *Practice of CSA* is shaped by <The Influence of Intermediaries in CSA>, such as platform operators and, in particular, enthusiastic intermediaries within them. At the same time, <The Weight of Decision-Making> varies depending on producers' own experiences and knowledge. Notably, producers who were initially hesitant to engage in CSA often engaged in <Meaning-Making That Sustains Practice>, for example, framing the activity as an opportunity to visit family when delivering shares. In addition, in the context of CSA's regular deliveries, producers addressed concerns about mismatches in production timing through <Coping with Production Uncertainty>, such as adjusting the quantity in subsequent deliveries or adding processed products. Through these [Internal Adjustments for Embarking on Practice], producers ultimately arrived at the *Practice of CSA*.

This process of adjustment is influenced by <Producers' Awareness and Value Understanding of CSA> as well as <The Continuity of Existing Marketing Channels> when producers compare CSA with their already established sales outlets, together constituting their [Understanding of CSA Ideals and Perceived Homogeneity with Other Sales Channels]. Producers who participate in CSA are often engaged in pesticide-free or reduced-pesticide cultivation, and consequently tend to practice small-scale, diversified farming. However, given that Japan's current food system is premised on large-scale monoculture production and the existence of stringent JAS organic certification standards, producers experience <Difficulty in Aligning with Existing Market Systems>. As a result, while they seek to establish farming practices less dependent on the conventional food system, a gap persists between <Self-Determination in Farm Management and Social Evaluation>.

The difficulty of integrating into existing food systems underscores the need to secure <Strategic Diversity of Marketing Channels>, making [Managerial Autonomy Under Market Constraints] a critical element for producers. The information and experience gained while exploring such farm management approaches further shape their [Understanding of CSA Ideals and Perceived Homogeneity with Other Sales Channels.] Once producers transition to the *Practice of CSA*, they engage in [Reality-Based Value Recognition.] This includes both <Perceived Value Through Practice> and the gap between their prior knowledge and value understanding of CSA. Such recognition encompasses not only social dimensions but also <Economic Significance and Challenges>, such as the implications of advance payment and the financial burdens inherent in practice.

This process of value recognition is closely linked to interactions with consumers. Producers perceive the <Characteristics of CSA Participant Segment> as representing a highly conscious group, thereby highlighting the [Specificity of CSA Participants] within the broader population of Japanese consumers. However, since there are also participants who have never engaged in face-to-face interactions or who are difficult to contact, a process of <Barriers and Efforts Toward Mutual Understanding Between Producers and Consumers in CSA> emerges. This process, in turn, influences producers' recognition of the value of CSA.

Producers recognize the <Weakness of Producer-to-Producer Exchange> with other local farmers or those engaged in CSA, although they also understand the busy schedules of fellow producers; consequently, little change occurs in these relationships. At the same time, they directly experience the <Changing Environment Surrounding Primary Industries>, including the rising interest of younger generations in agriculture, policy trends related to regional revitalization, and the marked decline in farmland due to aging farmers and farm abandonment. These [Challenges Surrounding Primary Industries] influence producers' managerial autonomy and reinforce the importance of interaction with consumers as a means to communicate and share these changes. For some producers, CSA thus becomes a space where such challenges and transformations of primary industries can be conveyed to consumers, thereby providing additional value.

Through this process of practice and efforts toward consumer understanding, the favorable relationships cultivated with certain participants or with partners such as cafés and restaurants serving as hubs function as a <Relational Ties that Encourage Continuation>, motivating producers to *Continuation of CSA*. However, <Transitional Collaborative Relationships Between Intermediaries and Producers>, shaped by the constraints and operational considerations of intermediaries, can at times exert a negative influence on producers' willingness to continue.

**Appendix B2 Storyline for consumers**

In the following presentation, categories are indicated as [Category], subcategories as <Subcategory>, and concepts as *concept*.

Consumers develop [Formation of Interest Through Environmental and Social Backgrounds] shaped by their upbringing and broader social trends, and they engage in various *Personal Concerns and Actions Regarding Environmental and Social Issues*. In the context of the [Limited Practice of CSA], consumers become aware of CSA while gathering information about environmental and social issues through <Diverse Information Channels>.

Through *Invitations from Acquaintances* encountered in the course of their engagements, consumers come to recognize CSA, and their *Empathy with the Concept of CSA*, together with <Influence from Other Purchasing Channels> such as subscription schemes and a <Proactive Orientation Toward Experiences and Connections> with farming, foster [From Initial Interest to the Formation of Willingness to Participate]. The cultivation of such willingness is shaped not only by *Personal Concerns and Actions Regarding Environmental and Social Issues* but also by <Transformation of Lifestyles>, such as seeking contact with farming or access to safe food as prompted by child-rearing, relocation, or the COVID-19 pandemic. With this initial interest and willingness, consumers move toward *Participation in CSA*.

This participation is further influenced by <Cooking Skills and Preferences for Utilizing CSA Vegetables>, formed through prior experiences with vegetable subscriptions and cooking, as well as by [Skills and Lifestyles that Influence the Ease of Participation], such as the ability to make time for share pick-up.

After engaging in CSA, consumers experience an [Enhanced Understanding of Production Through Direct Interaction] with producers. Through such direct interactions, they come to recognize the <Differences in the Positions of Producers and Consumers>, develop feelings of gratitude and respect, and deepen their understanding of production by gaining *Behind-the-Scenes Insights into Production and CSA* through <Building Relationships with Producers>.

Producers are involved not only in CSA but also in a range of other activities, the prior awareness of which influences *Participation in CSA*. These activities also bring about the

<Diversity and Richness of CSA-Related Experiences>, providing consumers with [Diverse Opportunities Accompanying Participation]. Moreover, the *Roles and Innovations of Intermediaries*, who connect producers and consumers, further contribute to this diversity of experiences. While these diverse opportunities enhance the [Realities Accompanying Participation], including <Participation Satisfaction and Gaps>, <Connections and Exchanges Among Consumers>, and <Emotional Gaps with Other Consumers>, they can, at times, also add to participants' burdens.

These experiences lead consumers to engage in <Comparison and Interaction with Other Activities> they had previously participated in, influencing both their participation in CSA and their involvement in other activities. Consumers also come to hold expectations for food security in times of crisis through <Building Relationships with Producers> and to develop a <Recognition of the Social Functions of Agriculture and CSA> through <Connections and Exchanges Among Consumers>.

Through CSA participation and assisting with farm work at CSA farms, consumers experience <Self-Growth and Future Expectations>. By considering their <Outlook Toward the Future> based on engagement with agriculture, they begin to form [Envisioned Ideals for the Future]. These ideals act upon their *Normative Consciousness Regarding Agriculture and Food in Society* and their <Meta-Reflection on Actions and Values>, leading to the [Formation and Renewal of Internal Behavioral Foundations]. Such internal foundations reduce barriers to <Participation and Continuation in CSA>, and consumers who have built positive relationships with producers and fellow consumers or who have gained satisfaction from participation are more likely to reach *Continuation of CSA*.